\documentclass[10pt,conference]{IEEEtran} 
\IEEEoverridecommandlockouts

\usepackage{url} 
\usepackage{framed}
\usepackage{mdframed}
\usepackage{listings}
\usepackage{multicol}
\usepackage{amssymb}
\usepackage{lipsum}
\usepackage{adjustbox}
\usepackage[font=bf,skip=\baselineskip]{caption}
\usepackage{tabularx}
\usepackage{seqsplit}
\usepackage{multirow}
\usepackage{booktabs}
\usepackage{graphicx}
\usepackage{hyperref}
\usepackage{subcaption}
\usepackage{makecell}
\usepackage{xcolor}
\usepackage[ruled,vlined,linesnumbered] {algorithm2e}
\usepackage{amsmath}
\usepackage{tikz}
\usepackage{tcolorbox}
\usepackage{threeparttable}
\usepackage{tablefootnote}
\usepackage{csquotes}
\usepackage{colortbl}
\usepackage{algorithm2e}
\usepackage{adjustbox}
\usepackage{balance}
\usepackage{dblfloatfix}
\usepackage{placeins}
\usepackage{afterpage}
\usepackage{amsthm}

\usepackage{breakurl} 
\usepackage{boldline}
 
\usepackage{pifont}% http://ctan.org/pkg/pifont
\usepackage{balance}
\let\labelindent\relax
\usepackage[ruled,vlined]{algorithm2e}
\usepackage{enumitem}
\usepackage[htt]{hyphenat}

\usepackage[normalem]{ulem}
\useunder{\uline}{\ul}{}

\usepackage{array}

\definecolor{customblue}{HTML}{006ca6}
\definecolor{customgreen}{HTML}{009264}
\definecolor{custombrown}{HTML}{ff3d00}
\AtEndPreamble{
\usepackage{hyperref}

    \hypersetup{
      colorlinks = true,
      linkcolor = customgreen,
      anchorcolor = purple,
      citecolor = customgreen,
      filecolor = purple,
      urlcolor = customgreen
    }
}

\newtheorem{definition}{Definition}

\newcommand{\alg}{\textsc{PESO}}
\newcommand{\tool}{\textsc{CodeMorph}}
\newcommand{\ea}{et~al.}

\begin{document}

\title{\tool{}: Mitigating Data Leakage in Large Language Model Assessment}

\author{
\IEEEauthorblockN{Hongzhou Rao\IEEEauthorrefmark{2}, Yanjie Zhao\IEEEauthorrefmark{2}\IEEEauthorrefmark{1}, Wenjie Zhu\IEEEauthorrefmark{2}, Ling Xiao\IEEEauthorrefmark{2}, Meizhen Wang\IEEEauthorrefmark{2}, and Haoyu Wang\IEEEauthorrefmark{2}}

\IEEEauthorblockA{\IEEEauthorrefmark{2}Huazhong University of Science and Technology, Wuhan, China \\
\{rhz, yanjie\_zhao, wjzhu, lingx, mzwang, haoyuwang\}@hust.edu.cn
}

\thanks{\IEEEauthorrefmark{1}Yanjie Zhao is the corresponding author (yanjie\_zhao@hust.edu.cn).}

\thanks{\IEEEauthorrefmark{2}The full name of the author's affiliation is Hubei Key Laboratory of Distributed System Security, Hubei Engineering Research Center on Big Data Security, School of Cyber Science and Engineering, Huazhong University of Science and Technology.}
}

\maketitle

\begin{abstract}

Concerns about benchmark leakage in large language models for code (Code LLMs) have raised issues of data contamination and inflated evaluation metrics. The diversity and inaccessibility of many training datasets make it difficult to prevent data leakage entirely, even with time lag strategies. Consequently, generating new datasets through code perturbation has become essential. However, existing methods often fail to produce complex and diverse variations, struggle with complex cross-file dependencies, and lack support for multiple programming languages, which limits their effectiveness in enhancing LLM evaluations for coding tasks.
To fill this gap, we propose \tool{}, an approach designed to support multiple programming languages while preserving cross-file dependencies to mitigate data leakage. \tool{} consists of two main components that work together to enhance the perturbation process. The first component employs 26 semantic-preserving transformation methods to iteratively perturb code, generating diverse variations while ensuring that the modified code remains compilable. 
The second component introduces a genetic algorithm-based selection algorithm, \alg{}, to identify the more effective perturbation method for each iteration by targeting lower similarity scores between the perturbed and original code, thereby enhancing overall perturbation effectiveness.
Experimental results demonstrate that after applying \tool{}, the accuracy of the LLM on code completion tasks across five programming languages decreased by an average of 24.67\%, with Python showing the most significant reduction at 45\%. The similarity score of code optimized by \alg{} is, on average, 7.01\% lower than that of randomly perturbed code, peaking at a reduction of 42.86\%. Additionally, overall accuracy dropped by an average of 15\%, with a maximum decrease of 25\%. These findings indicate that \tool{} effectively reduces data contamination while \alg{} optimizes perturbation combinations for code.

\end{abstract}

\section{Introduction}

As Large Language Models for Code (Code LLMs) continue to advance~\cite{lozhkov2024starcoder,guo2024deepseek,roziere2023code}, researchers test these LLMs by having them complete a series of code tasks on evaluation datasets, scoring their performance based on task completion and evaluation metrics to represent the LLM's capabilities. For example, Meta AI evaluated Code Llama~\cite{roziere2023code} on HumanEval~\cite{feng2024complexcodeeval} and MBPP~\cite{austin2021program}, demonstrating its performance superiority over some smaller models such as PaLM-Coder~\cite{anil2023palm} and OpenAI's code-cushman-001. However, Yang \ea{}~\cite{yang2023rethinking} and Riddell \ea{}~\cite{riddell2024quantifying} have demonstrated that both HumanEval and MBPP datasets exhibit varying degrees of data leakage. \textbf{Data leakage in LLM assessment occurs when the LLM encounters portions of the evaluation dataset during its training phase}, which can lead to artificially inflated performance scores~\cite{yang2023rethinking}. This phenomenon raises concerns regarding the reliability of LLM evaluation results and may undermine their applicability in real-world scenarios~\cite{shi2023detecting,zhou2023don,xu2024benchmark}. Nevertheless, given the vast scale and diversity of knowledge included in LLM training corpora, designing a completely contamination-free benchmark remains an extremely challenging task~\cite{yan2023codescope}.
% This phenomenon has raised doubts about the authenticity of LLM performance and may impact the application of LLMs in real-world scenarios~\cite{shi2023detecting,zhou2023don,xu2024benchmark}. However, due to the vast scale and diversity of knowledge in LLM training corpora, creating a contamination-free benchmark is extremely challenging~\cite{yan2023codescope}.

Based on the concerns mentioned above, researchers have begun exploring solutions to data contamination. Cao \ea{}~\cite{cao2024concerned} found that data perturbation could help mitigate the effects of data contamination. Data perturbation is a technique that involves applying a series of modifications to the original data so that the perturbed data differs from the original. For code, the goal is to ensure that the perturbed code remains both functional and consistent with its original functionality, adhering to strict logical and syntactical requirements to achieve semantic-preserving transformations defined as \textbf{Definition} \autoref{def:code trans}. Fortunately, semantics-preserving transformations for code have a well-established research history, with active studies continuing in recent years. For example, Sun \ea{}~\cite{sun2023codemark} utilize methods such as syntactic sugar and default parameters to make watermarking for code. Zhang \ea{}~\cite{zhang2023statfier} generates program variants by using methods like for/while loops transformation and dead code insertion. 

However, since benchmarks for evaluating LLMs often involve multiple languages and real-world code, the following challenges are presented for methods intended to perturb code for LLM evaluation.
First, there is a need for \textbf{more complex perturbations}. While researchers like Sun et al.~\cite{sun2023codemark} and Zhang et al.~\cite{zhang2023statfier} have utilized common semantic-preserving transformations, more sophisticated methods are necessary to minimize data contamination.
Another challenge is \textbf{ensuring applicability across multiple programming languages}. Current tools such as \texttt{CodART}~\cite{CodART}, \texttt{rewrite}~\cite{rewrite}, and \texttt{rope}~\cite{rope} are limited to specific languages, primarily Java or Python. Although \texttt{Comby}~\cite{comby} supports multiple languages, its dependence on manually defined matching patterns restricts automatic refactoring capabilities.
Additionally, \textbf{managing cross-file dependencies} is crucial. Most benchmarks rely on real-world data with complex interdependencies, which must be preserved during perturbation to ensure the functionality of the modified code.
Finally, any perturbation must \textbf{guarantee correct execution}. The perturbed code should compile and run successfully, preserving the original functionality to provide a reliable evaluation of LLMs.

To address these challenges, we propose \tool{}, an approach that supports multiple programming languages for effective code perturbation, including C/C++, Go, Python, Rust, and Java. \tool{} consists of two main components. The first component performs two key functions: code perturbation and code verification. We introduce 26 types of semantics-preserving code transformations that can be executed by LLMs. By combining different perturbation methods, LLMs can achieve more complex modifications that would be challenging to accomplish with static tools. After perturbation, the code undergoes verification to ensure successful compilation and semantic equivalence with the original code. 

The second component is designed to identify more effective perturbations to enhance their overall efficacy. To achieve this, we propose a genetic algorithm-based selection algorithm, \alg{}. \alg{} is composed of two key elements: measuring the perturbation effectiveness of each method category and selecting perturbation methods based on their effectiveness. First, \alg{} employs a similarity score to quantify the effectiveness of each method—where a lower similarity score indicates a more effective perturbation. The similarity score is updated iteratively. Subsequently, using Boltzmann selection~\cite{kirkpatrick1983optimization}, a more effective perturbation method is chosen for each iteration. This approach ensures that perturbations with higher effectiveness are retained with a high probability.

We evaluate \tool{} through five experimental groups and two ablation test groups using random perturbations, totaling 720 code completion tasks, addressing two research questions (RQs). First, we construct three types of code completion tasks for each experimental group and record their accuracy (RQ1). Additionally, we conduct two ablation experiments to assess the effectiveness of \alg{} (RQ2). The results demonstrate that \tool{} reduces the accuracy of code completion tasks for perturbed code. Furthermore, code perturbed with \alg{} exhibits lower code similarity and accuracy in completion tasks compared to code perturbed by random methods. This indicates that \tool{} effectively mitigates data leakage, while \alg{} enhances the impact of code perturbations by selecting perturbation methods.

Overall, our contributions can be summarized as follows:

\begin{itemize}
    \item We propose an approach named \tool{} to mitigate data leakage in LLM assessment. \tool{} includes 26 types of semantically equivalent transformations, enables complex transformations that are difficult to achieve with static tools, supports perturbations across multiple languages, and offers good scalability. Additionally, \tool{} preserves repos-level dependencies, ensuring that the perturbed code retains its original functionality and compiles correctly.
    
    \item We introduce a \textbf{p}erturbation m\textbf{e}thod \textbf{s}election \textbf{o}ptimization algorithm (\alg{}) based on a genetic algorithm. \alg{} identifies more effective perturbation methods to enhance effectiveness. Additionally, it avoids restricting the search to a limited retrieval space, reducing the risk of converging on local optima and enabling a thorough exploration of the perturbation method space to ensure greater diversity in method selection.
    
    \item Experimental results indicate that \tool{} reduces code completion accuracy of the LLM by an average of 24.67\%, with a maximum reduction of 45\%. Additionally, compared to random method selection, \alg{} achieves an average reduction in similarity score of 7.01\%, with a maximum reduction of 42.86\%, and an average reduction in code completion accuracy of 15\%, with a maximum reduction of 25\%.
\end{itemize}

\noindent\textbf{Artifact Availability.} Our artifact is available at \url{https://github.com/security-pride/CodeMorph}.

\begin{table}[ht!]
\center
\caption{Perturbation methods and their corresponding categories}
\label{tab:all-methods}
\resizebox{\linewidth}{!}{%
\begin{tabular}{lp{0.4\linewidth}p{0.5\linewidth}}
\hline
\textbf{Category}                         & \multicolumn{1}{c}{\textbf{Methods}}                                                                                                                                                            & \multicolumn{1}{c}{\textbf{Description}}                                                                                    \\ \hline
\multirow{11}{*}{\begin{tabular}[c]{@{}l@{}}Basic\\ Methods\end{tabular}}         & Add exception*                                                                                                                                                                         & Add exception handling blocks                                                                                      \\ \cline{2-3} 
                                       & Add arguments~\cite{dong2024effectiveness,sun2023codemark}                                                                                     & Add arguments to the function                                                                                      \\ \cline{2-3} 
                                       & Change statement order~\cite{zhang2023challenging}                                                                                                                     & \begin{tabular}[c]{@{}l@{}}Change the order of two adjacent statements \\ that do not share any variables.\end{tabular} \\ \cline{2-3} 
                                       & Check arguments~\cite{dong2024effectiveness}                                                                                                                          & Check if the arguments are none                                                                                    \\ \cline{2-3} 
                                       & Insert junk function*                                                                                                                                                                  & Insert junk functions that won't be called                                                                         \\ \cline{2-3} 
                                       & Insert junk loop~\cite{zhang2023statfier,zhang2023challenging,dong2024effectiveness}                                    & Insert junk loops that won't be executed                                                                           \\ \cline{2-3} 
                                       & Insert variables\cite{dong2024effectiveness}                                                                                                                          & Insert variables that won't be used                                                                                \\ \cline{2-3} 
                                       & Move assignments~\cite{zhang2023statfier}                                                                                                                             & \begin{tabular}[c]{@{}l@{}}Move assignments if a variable is assigned\\ directly\end{tabular}                                                                \\ \cline{2-3} 
                                       & Statement wrapping~\cite{zhang2023statfier}                                                                                                                           & \begin{tabular}[c]{@{}l@{}}Wrap the statements with \texttt{if} or \texttt{for} \\ statement\end{tabular}                                                                \\ \cline{2-3} 
                                       & Function rename~\cite{zhang2023challenging,yu2022data,dong2024effectiveness}                                            & Change the function name                                                                                           \\ \cline{2-3} 
                                       & Variables rename~\cite{zhang2023challenging,yu2022data,dong2024effectiveness}                                           & Change the variables name                                                                                          \\ \hline
\multirow{6}{*}{\begin{tabular}[c]{@{}l@{}}Condition\\ Methods\end{tabular}}     & Add conditon*                                                                                                                                                                          & \begin{tabular}[c]{@{}l@{}}Improve condition statements by adding \\\texttt{else} clauses\end{tabular}                                                                \\ \cline{2-3} 
                                       & Div if else~\cite{zhang2023challenging}                                                                                                                                & \begin{tabular}[c]{@{}l@{}}Divide the \texttt{if else-if else} into \texttt{if}\\  \texttt{else if else}\end{tabular}                                                           \\ \cline{2-3} 
                                       & Div composed if~\cite{yu2022data}                                                                                                                                      & \begin{tabular}[c]{@{}l@{}}Divide the composed \texttt{if} statement into  single\end{tabular}                                                              \\ \cline{2-3} 
                                       & If-continue to if-else~\cite{yu2022data}                                                                                                                               & \begin{tabular}[c]{@{}l@{}}Transform \texttt{if-continue} statement to \\ \texttt{if-else} statement\end{tabular}                                                     \\ \cline{2-3} 
                                       & If to switch/match~\cite{zhang2023challenging,yu2022data}                                                                                      & \begin{tabular}[c]{@{}l@{}}Transform \texttt{if} statement to \texttt{switch/match}\\  statement\end{tabular}                                                          \\ \cline{2-3} 
                                       & Switch/match to if~\cite{zhang2023challenging,yu2022data}                                                                                       & \begin{tabular}[c]{@{}l@{}}Transform \texttt{switch/match} statement to  \texttt{if}  \\statement\end{tabular}                                                          \\ \hline
\multirow{2}{*}{\begin{tabular}[c]{@{}l@{}}Loop\\ Methods\end{tabular}}          & Div loop*                                                                                                                                                                              & Divide a loop into several loops                                                                                   \\ \cline{2-3} 
                                       & \begin{tabular}[c]{@{}l@{}}For/while transformation~\cite{zhang2023statfier},\\ \cite{zhang2023challenging,yu2022data}\end{tabular}                                        &\begin{tabular}[c]{@{}l@{}}Transform \texttt{for/while} loop to  \texttt{while/for} \\ loop\end{tabular}                                                            \\ \hline
\multirow{2}{*}{\begin{tabular}[c]{@{}l@{}}Logic\\ Methods\end{tabular}}         & Equi boolean logic*                                                                                                                                                                    & Transform boolean logic equivalently                                                                \\ \cline{2-3} 
                                       & \begin{tabular}[c]{@{}l@{}}Swap boolean expression~\cite{zhang2023statfier}, \\ \cite{zhang2023challenging,yu2022data,sun2023codemark}\end{tabular} & Swap the sides of the boolean expression                                                                           \\ \hline
\multirow{2}{*}{\begin{tabular}[c]{@{}l@{}}Decomposition\\ Methods\end{tabular}} & Extract if*                                                                                                                                                                     & Extract method from \texttt{if} statements                                                                               \\ \cline{2-3} 
                                       & Extract arithmetic*                                                                                                                                                             & Extract method from arithmetic statements                                                                                     \\ \hline
\multirow{3}{*}{\begin{tabular}[c]{@{}l@{}}Arithmetic\\ Methods\end{tabular}}    & \begin{tabular}[c]{@{}l@{}}Equi arithmetic expression~\cite{zhang2023statfier},\\ \cite{dong2024effectiveness}\end{tabular}                                                                       & \begin{tabular}[c]{@{}l@{}}Change the arithmetic computation or \\arithmetic assignment equivalently\end{tabular} \\ \cline{2-3} 
                                       & Expression div~\cite{yu2022data}                                                                                                                                       & \begin{tabular}[c]{@{}l@{}}Divde the long expression to several small\\  expressions\end{tabular}                                                             \\ \cline{2-3} 
                                       & Modify operations~\cite{zhang2023challenging,sun2023codemark}                                                                                  & \begin{tabular}[c]{@{}l@{}}Modify the compound assignment operations \\ such as \texttt{a += b} to \texttt{a = a + b}\end{tabular}                                                                         \\ \hline
\end{tabular}%
}

\vspace{0.1cm}

\parbox{\linewidth}{\footnotesize \raggedright Note: Methods marked with * are proposed by us and are not listed in \cite{dong2024effectiveness, zhang2023statfier, zhang2023challenging, yu2022data, sun2023codemark}.}
\end{table}

\section{Preliminary Study}\label{sec:pre-study}

Given the impressive performance of LLMs in code-related tasks~\cite{hou2023large}, we conduct a preliminary study to explore their potential for code perturbation.
To ensure the rigor of our experiments, we require an LLM capable of executing code tasks and a dataset that guarantees data leakage for this LLM. The StarCoder LLM~\cite{starcoder} has publicly released its training dataset, The Stack dataset~\cite{thestack}, which is well-documented and confirmed to be ``contaminated''. This makes it suitable for our experimental needs. Therefore, \textbf{we have selected StarCoder as the LLM to be evaluated and chosen code samples from The Stack dataset to construct our test samples}.

Before applying code perturbations, we first selected 26 semantic-preserving code transformation methods. Semantic-preserving code transformations in here are defined as follows:

\begin{definition}[Semantic-preserving transformations]\label{def:code trans}
Let $P$ and $P'$ be two programs in a programming language $\mathcal{L}$, with the same input space $\mathcal{I}$ and output space $\mathcal{O}$. $P'$ is called a \textbf{semantics-preserving transformation} of $P$ if for any input $x \in \mathcal{I}$, the following condition holds:
\[f_P(x) = f_{P'}(x, \text{extra\_params})\]
\noindent where: $f_P : \mathcal{I} \to \mathcal{O}$ and $f_{P'} : \mathcal{I} \times \text{extra\_params} \to \mathcal{O}$ are the input-output mapping functions of programs $P$ and $P'$, \text{extra\_params} denotes the set of additional parameters that do not influence the output of $P'$ under any circumstances.
    % , denoted as:

    % \[P \overset{\text{semantics-preserving}}{\longrightarrow} P'\]

\end{definition}

After selecting the transformation methods, we chose one Python code sample and one Go code sample from The Stack. We then applied 26 distinct perturbation methods to each sample, as detailed in \autoref{tab:all-methods}, generating a total of 52 perturbed code samples. Subsequently, we conducted code completion tasks on these 52 samples using StarCoder. The results revealed an accuracy of 94.3\% across the tasks, demonstrating that a single perturbation is \textbf{not effective} in impacting performance.

\begin{figure}[htbp]  
    \centering     
    \includegraphics[width=0.9\linewidth]{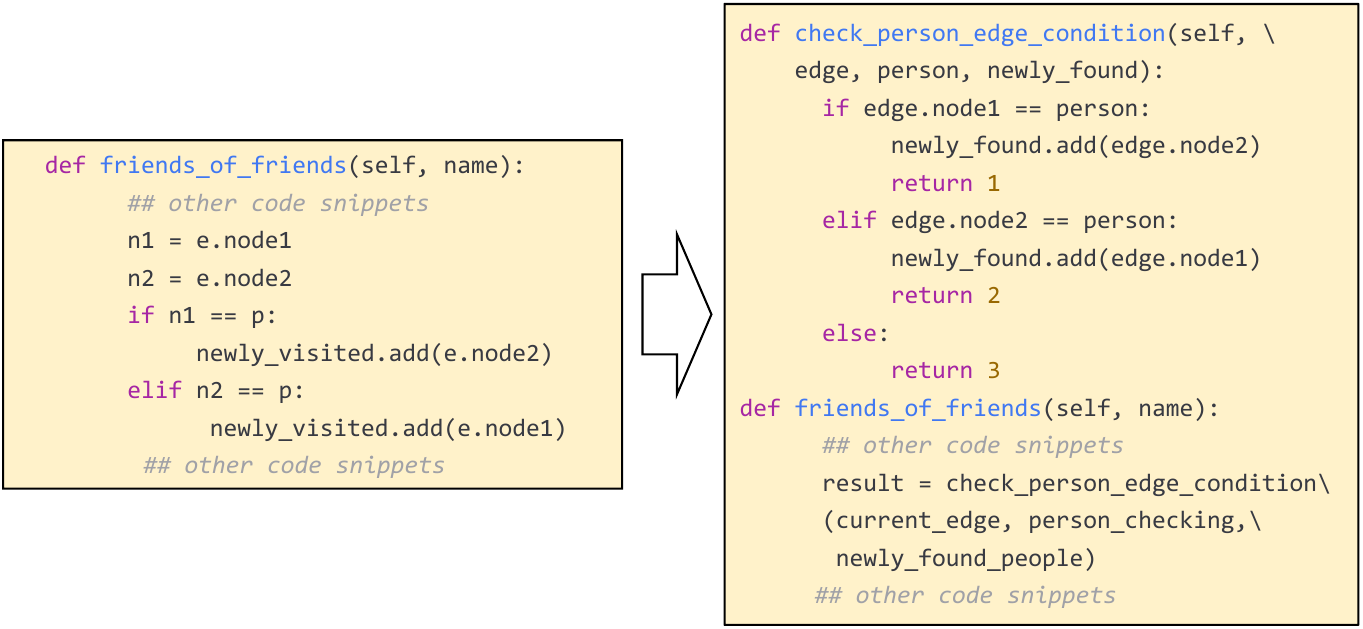}  
    \caption{The left side shows the original code, which is transformed into the perturbed code on the right side using the \textit{Extract If} method}
    \label{fig:code-perturbation-example}

\end{figure}

Since single perturbations have limited effectiveness, we selected additional samples for comprehensive experiments to investigate whether combinations of multiple perturbation methods can produce more effective perturbations. Specifically, we selected an additional 10 Python codes and applied multiple perturbations to each code, with each perturbation method randomly selected from the 26 methods shown in \autoref{tab:all-methods}.

The first phenomenon we observed is that the LLM can transform the original code into a more \textbf{complex} version. As shown in \autoref{fig:code-perturbation-example}, the LLM uses the \emph{Extract if} method to extract \texttt{e.node1} and \texttt{e.node2} as conditional checks, omitting the intermediate variables \texttt{n1} and \texttt{n2} and encapsulating them into a function that returns the evaluation result. This indicates that the LLM perturbs the code based on the \texttt{if} conditions, achieving a complex perturbation. In contrast, static tools can only perturb code according to predefined patterns, making it challenging to apply context-specific perturbations. As a result, the effectiveness of static tools is limited.

\begin{figure}[ht!]  
    \centering     
    \includegraphics[width=\linewidth]{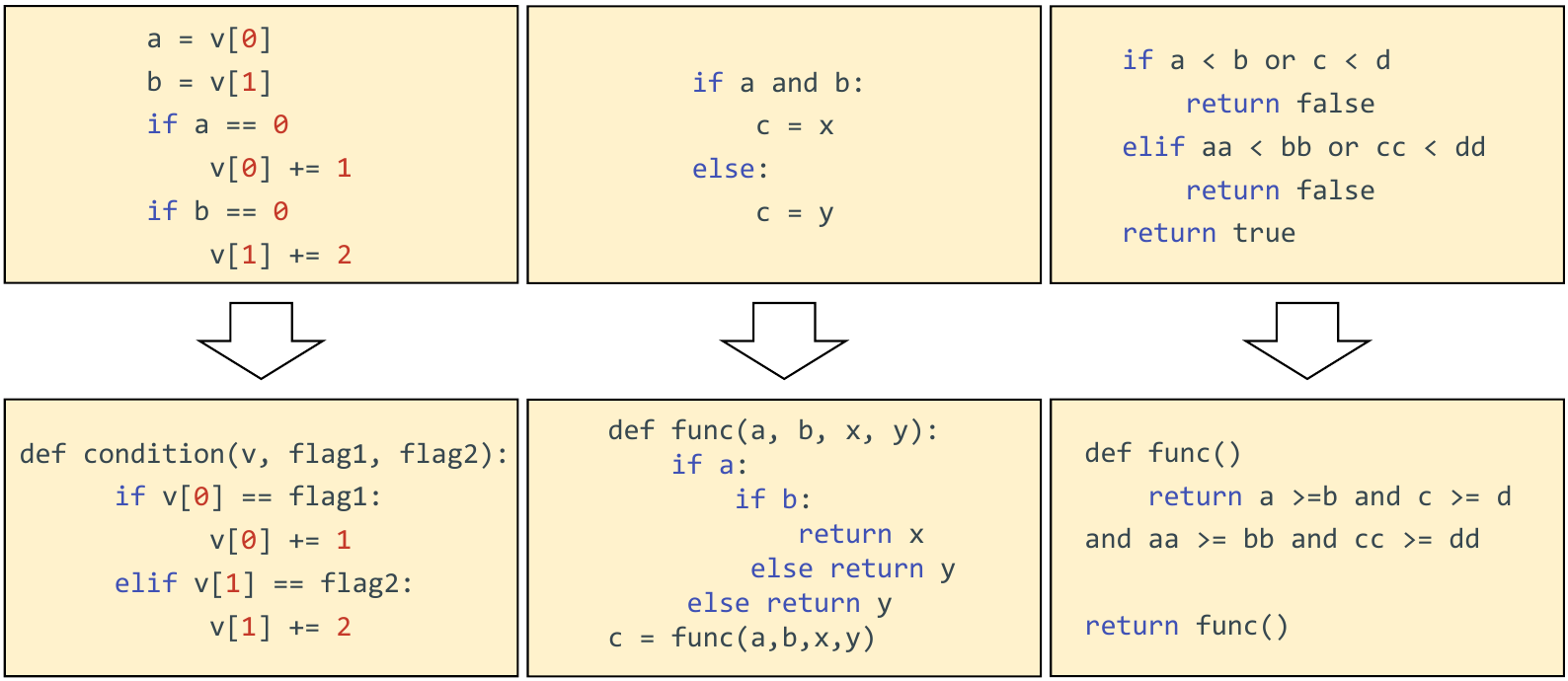}  
    \caption{The code above is the original, while the code below shows the perturbed versions. The original code on the left only uses the \textit{Extract if} method. The code in the middle applies both the \textit{Extract if} method and the \textit{Div if else} method. The code on the right applies the \textit{Extract if} method along with the \textit{Equi Boolean Logic} method}
    \label{fig:code_trans_sample} 
\end{figure}

The second phenomenon we observed is that the LLM can produce \textbf{diverse} perturbations. \autoref{fig:code_trans_sample} shows three simple examples, each using the \textit{Extract if} method. The example on the left is similar to \autoref{fig:code-perturbation-example}, where the \texttt{if} statement is encapsulated into a function. The middle example differs slightly in that the encapsulated function returns only \texttt{x} or \texttt{y} after separating the \texttt{if} statement. Like the middle example, the example on the right returns a single value, but it consolidates the \texttt{if} statements to make the code more concise. This shows that LLMs can produce diverse perturbation effects using the same method, depending on different code structures, highlighting their versatility in code perturbation.

\begin{figure}[htbp]  
    \centering     
    \includegraphics[width=\linewidth]{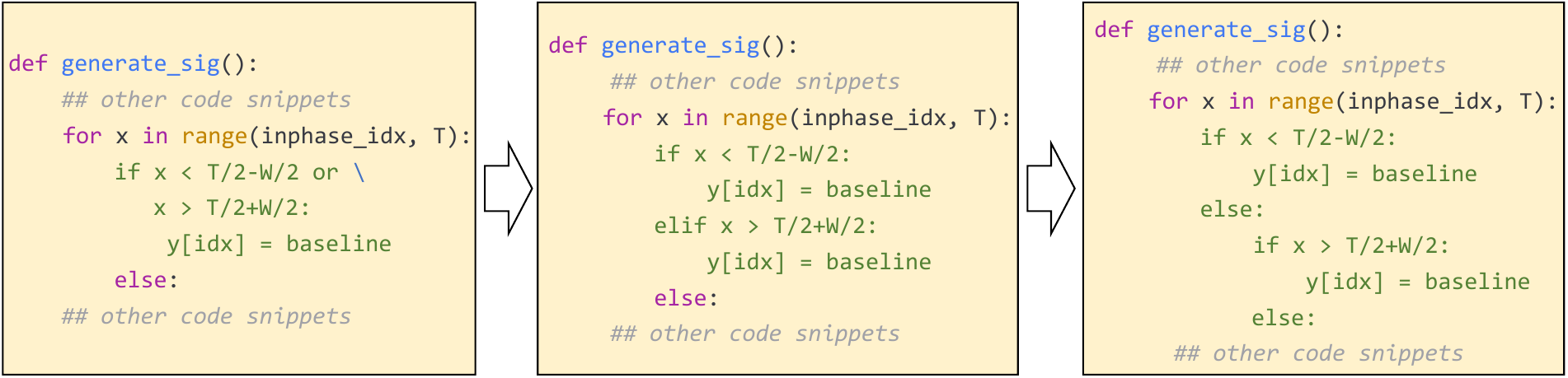}  
    \caption{The left example shows the original code. The code in the middle is the original code perturbed with the \emph{Div composed if} method, and the code on the right is the code further perturbed from the middle version using the \emph{Div if else} method. Perturbations are highlighted in green}
    \label{fig:order_of_perturbation} 
\end{figure}

The third phenomenon we observed is that different \textbf{combinations} and \textbf{orders} of perturbations affect the outcome of code perturbations. As shown in \autoref{fig:order_of_perturbation}, after applying the \emph{Div composed if} method to the original code, the middle perturbed code includes an added \texttt{elif} statement, which then satisfies the conditions for further perturbation using the \emph{Div if else} method. Had we initially applied the \emph{Div if else} method to the original code, it would have been unsuitable, as the original code lacks \texttt{elif} statements and is better suited to the \emph{Div composed if} method. Similarly, methods like \emph{Move assignment} would not be effective for the original code on the left, demonstrating that the combination of \emph{Div composed if} followed by \emph{Div if else} is more appropriate for perturbing the original code. This illustrates the importance of perturbation combinations and order on the outcome, underscoring the need to determine which perturbation method is more appropriate for the current code structure.

Based on the above conclusions, we propose an automatic code perturbation method using LLMs. Several considerations for choosing LLMs to perturb code are summarized as follows:
\textbf{(1) Scalability:} Unlike static tools, LLMs possess extensive knowledge of code and are highly scalable. They can be easily adapted to support additional programming languages through tailored prompt design, effectively addressing the challenge of \textbf{applicability across multiple programming languages}.
 \textbf{(2) Complex perturbation:} By leveraging their understanding of code and creative capabilities, LLMs can combine multiple simple perturbation methods to generate complex code modifications—something that is difficult to achieve with static tools. This capability effectively addresses the challenge of \textbf{creating more sophisticated perturbations}. \textbf{(3) Diverse perturbation:} In contrast to static tools limited to preset patterns, LLMs can adapt to modified code structures to align with specific perturbation methods, resulting in a wider variety of perturbation types.

However, there are several challenges associated with using LLMs for code perturbation: \textbf{(1)~Prompt design:} Guiding LLMs for code perturbation involves two major difficulties. First, LLMs have limitations in code reasoning, particularly when generating solutions for real-world problems~\cite{liu2024codemind}. Therefore, designing precise prompts to instruct LLMs on how to transform code is crucial. Second, to encourage LLMs to produce more diverse perturbations, it is necessary to abstract the code perturbation format to prevent them from only implementing a limited set of specific patterns. To address this, we propose a prompt synthesis approach. \textbf{(2)~Correct perturbation:} Due to the hallucinations of LLMs, the generated code may be uncompilable or not semantically equivalent to the original. Additionally, some repository-level benchmarks, such as CoderEval~\cite{yu2024codereval}, ComplexCodeEval~\cite{feng2024complexcodeeval}, and CrossCodeEval~\cite{ding2024crosscodeeval}, include extensive code dependencies. Disrupting these dependencies can cause the code to fail to compile or run correctly. To address these issues, we ensure that no dependencies are modified during perturbation and develop a multi-language execution engine to compile the generated code. Furthermore, previous studies~\cite{zhu2024dynamic,weyssow2024codeultrafeedback} have demonstrated the reliability of using LLMs as evaluators. Therefore, we employ three LLMs to assess whether the generated code is semantically consistent with the original, effectively tackling the challenges of \textbf{managing cross-file dependencies} and \textbf{ensuring correct execution}.
\textbf{(3)~Method selection:} As previously mentioned, the combination and order of perturbation methods significantly impact code perturbation. The challenge lies in discovering more effective perturbation methods. To address this issue, we propose a novel optimization algorithm based on genetic algorithms, \alg{}, to intelligently select the most effective perturbation method for each iteration, thereby generating more effective and diverse combinations and orders of perturbation methods for each code.

\begin{figure*}[htbp]  
    \centering     
    \includegraphics[width=\linewidth]{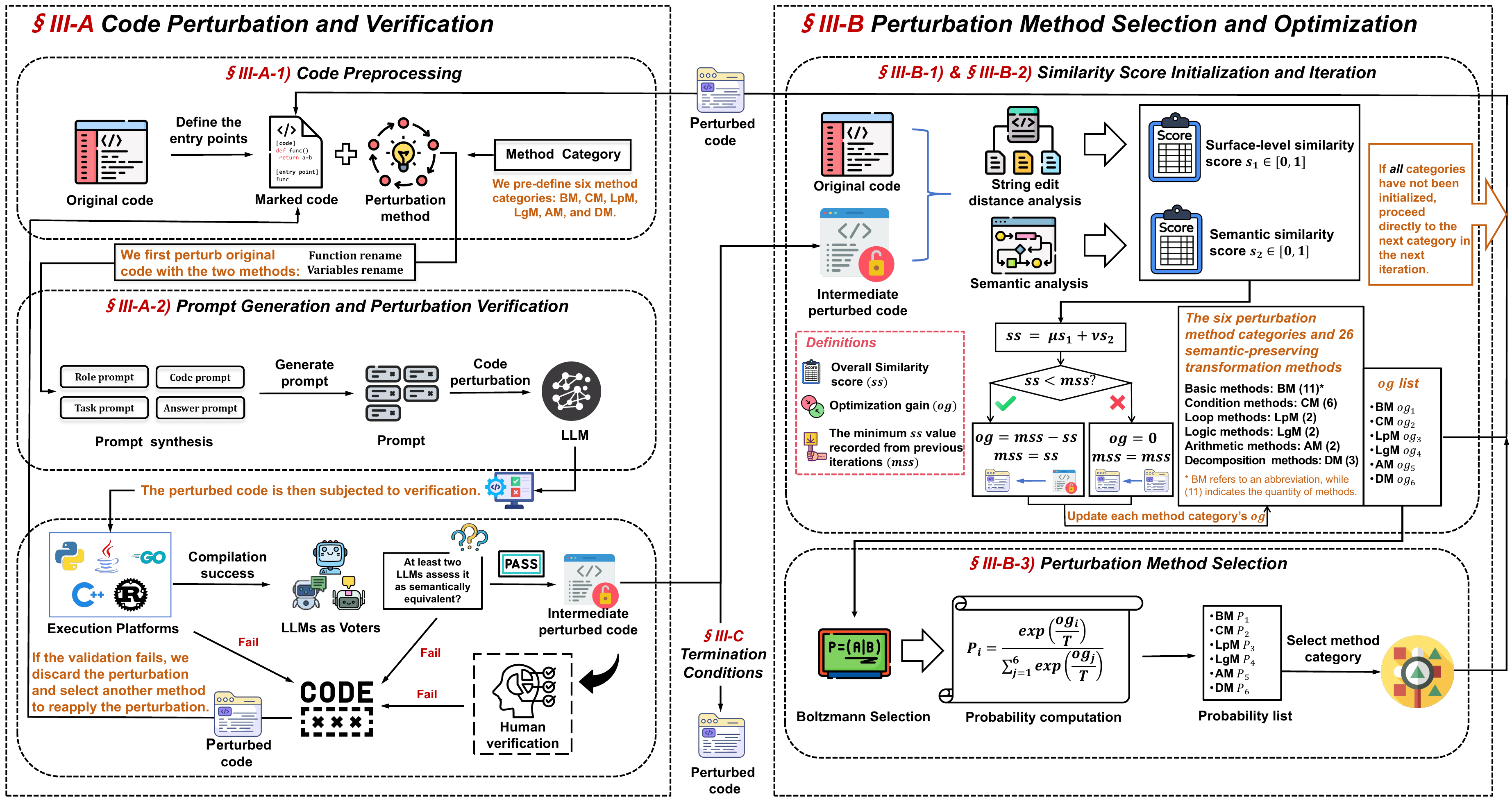}  
    \caption{The workflow of \tool{}}  
    \label{fig:framework}  
\end{figure*}

\section{Methodology}
\label{sec:approach}

\autoref{fig:framework} illustrates the framework of \tool{}, which consists of two components: code perturbation and verification, and perturbation method selection and optimization. For clarity, we have provided a description of the notations in \autoref{tab:notation-descri}.

\begin{table}[]
\caption{Notation description for \tool{}}
\label{tab:notation-descri}
\resizebox{\linewidth}{!}{%
\begin{tabular}{@{}ll@{}}
\toprule
\textbf{Notation} & \textbf{Description}                                                                                                                                                                                                    \\ \midrule
$c_{ori}$         & Original code that does not been perturbed                                                                                                                                                                              \\ \midrule
$c_{inter}$       & The intermediate perturbed code that has not been verified in the current iteration                                                                                                                                     \\ \midrule
$c_{pert}$        & The perturbed code                                                                                                                                                                                                      \\ \midrule
$ss$              & \begin{tabular}[c]{@{}l@{}}The overall similarity score, which measures the similarity between the intermediate \\ perturbed code $c_{inter}$ and the original code $c_{ori}$, composed of $s_1$ and $s_2$\end{tabular} \\ \midrule
$s_1$             & The surface-level similarity score                                                                                                                                                                                      \\ \midrule
$s_2$             & The semantic similarity score                                                                                                                                                                                           \\ \midrule
$mss$             & Minimum $ss$ value, which records the minimum $ss$ from previous iterations                                                                                                                                             \\ \midrule
$og$              & \begin{tabular}[c]{@{}l@{}}Optimization gain, representing the improvement achieved by the perturbed\\ code in the current iteration\end{tabular}                                                                       \\ \midrule
$og_i$            & The optimization gain of $mc_i$                                                                                                                                                                                                      \\ \midrule
$OG$              & A list containing all $og_i$ values                                                                                                                                                                                     \\ \midrule
$pm$              & The perturbation method selected in the current iteration                                                                                                                                                               \\ \midrule
$mc$              & The perturbation method category selected in the current iteration                                                                                                                                                      \\ \midrule
$mc_i$            & The $i^{\text{th}}$ perturbation method category in $MC$                                                                                                                                                                \\ \midrule
$MC$              & A list of all perturbation method categories                                                                                                                                                                            \\ \bottomrule
\end{tabular}%
}

\end{table}

At the start of code perturbation, we initialize $c_{pert}$ as $c_{ori}$ and always apply perturbations to $c_{pert}$, while keeping $c_{ori}$ unchanged for subsequent similarity calculations. To ensure that the code differs from $c_{ori}$ from the outset, we \textbf{heuristically} choose the initial perturbation methods as \emph{Function rename} and \emph{Variable rename} in sequence, rather than selecting randomly. 

We now introduce the overall process of \tool{}.
Initially, we generate $c_{inter}$ using the perturbation method \emph{Function rename}. We then verify whether $c_{inter}$ can be compiled and whether it is semantically equivalent to $c_{pert}$. If the code fails verification, we discard this perturbation and select a new perturbation method for the next iteration. Conversely, if the verification succeeds, we update $c_{pert}$ to $c_{inter}$. Note that in this phase, $c_{pert}$ does not enter the \textit{perturbation method selection and optimization} phase. Next, \tool{} applies the \emph{Variable rename} perturbation method to $c_{pert}$ and generates $c_{inter}$, then repeats the verification process.

Following initialization, each iteration of \tool{} progresses through four stages: perturbation, verification, similarity score iteration, and perturbation method selection, based on each category’s optimization gain $og_i$ (detailed in \autoref{sec:method/init}). In the $k^{\text{th}}$ iteration, \tool{} first perturbs $c_{pert}$ using the perturbation method $pm$.If the generated code $c_{inter}$ fails verification, the optimization gain $og$ is set to 0, and $OG$ is updated with this value as described in lines 14–15 of \autoref{alg:optimization-algorithm}. Meanwhile, $c_{pert}$ remains unchanged, leading directly to the next method selection. If $c_{inter}$ passes verification, \tool{} computes the overall similarity score $ss$.

If $ss$ is not greater than $mss$, we update $c_{pert}$ to $c_{inter}$, set $og = mss - ss$, and assign $ss$ to $mss$. Conversely, if $ss > mss$, we set $og = 0$, while leaving $mss$ and $c_{pert}$ unchanged. Following this, $OG$ is updated with $og$ as described in lines 14–15 of \autoref{alg:optimization-algorithm}. This structured approach ensures that each perturbation iteration consistently reduces code similarity. 

Finally, \tool{} selects the category $mc$ according to $OG$ and randomly chooses the perturbation method $pm$ from $mc$ for the next iteration.
The iteration continues until either the iteration limit is reached or the minimum similarity score threshold $ss_{threshold}$ is achieved, as detailed in \autoref{sec:termination}.

\begin{algorithm}[htbp]
    \footnotesize
    \caption{Optimization Algorithm for Perturbation Method Selection}
    \label{alg:optimization-algorithm} 
    \KwIn{
          Original code $c_{ori}$,\\
          Perturbed code $c_{pert}$,\\
          Selected perturbation methods $pm$,\\
          Optimization gain list $OG[og_1 \dots og_n]$, \\
          Method categories list $MC[mc_1 \dots mc_n]$} 
    \KwOut{Perturbed code $c_{pert}$}

    \SetKwFunction{FSim}{GetSimilarity}
    \SetKwFunction{Gindex}{GetIndex}
    \SetKwFunction{FMethod}{MethodSelection}
    \SetKwFunction{Pert}{Perturbation}
    \SetKwFunction{Init}{Initialization}

    $ss, mss \gets 1, 1$\;
    $OG, c_{pert} \gets \Init(MC, c_{pert})$\;\vspace{6pt}
    
    \For{$iter \gets 1$ \KwTo $maxIter$ \textbf{and} $ss < ss_{threshold}$}{
        
        $c_{inter} \gets \Pert(pm, c_{pert})$\;
        $iter \gets iter + 1$\;

        $s_1, s_2 \gets \FSim(c_{ori}, c_{pert})$\;

        $ss \gets \mu s_1 + \nu s_2$\;

        \If{$mss \ge ss$}{
            $og \gets mss - ss$\;
            $mss \gets ss$\;
            $c_{pert} \gets c_{inter}$\;
        }
        \Else{
            $og \gets 0$\;
        }

        $Index \gets \Gindex(mc_i, MC)$\;
        $OG[Index] \gets og$\;
        
        \For{$i \gets 1$ \KwTo $n$}{
            $P[i] \gets \frac{e^{og_i / T}}{\sum_{j=1}^{n} e^{og_j / T}}$\;
        }
    
        $MIndex \gets \text{randomly select according to } P$\;
        $ pm \overset{\text{randomly}}{\xleftarrow{\hspace{0.7cm}}} MC[MIndex] $
    }

    \Return $c_{pert}$\;
\end{algorithm}

\subsection{Code perturbation and verification}
\label{sec:3.1}

\subsubsection{Code preprocessing}

As shown in \autoref{tab:all-methods}, there are a total of 26 perturbation methods. Among these, the methods \emph{Add exception}, \emph{Add condition}, and \emph{Div loop} are specifically designed to enhance code robustness. These methods implement exception handling in error-prone areas, add missing conditions in \texttt{if} statements, and divide large loops into smaller loops to prevent memory crashes. Additionally, we introduce two code segmentation methods: \emph{Extract if} method and \emph{Extract arithmetic} method. These methods significantly alter the code structure, thereby reducing similarity scores.

To enable the method selection process (see \autoref{sec:method selection}) to better tailor appropriate methods to different code snippets, these methods have been classified into six categories. Among them, \emph{\textbf{Basic Methods}} are general-purpose and do not specifically target control flow or calculations, making them applicable to most code. \emph{\textbf{Condition Methods}} focus on conditional statements within the code, while methods in \emph{\textbf{Loop Methods}} and \emph{\textbf{Logic Methods}} perform semantics-preserving transformations on loops and logical conditions, respectively. \emph{\textbf{Decomposition Methods}} include two code segmentation methods \emph{Extract if} method and \emph{Extract arithmetic} method. \emph{\textbf{Arithmetic Methods}} contain methods effective for code with calculations. As noted previously, combining different methods can produce more complex perturbations. This underscores the importance of the \alg{} approach proposed in \autoref{sec:method selection}.

\subsubsection{Prompt generation and perturbation verification}\label{sec:promt-generation}

After selecting the perturbation method, a prompt must be provided to the LLM, which we refer to as the perturbation LLM, to execute the perturbation. To ensure that the perturbation LLM performs accurate perturbations and generates diverse code, we designed task-specific prompts. Each prompt consists of the following components:
\begin{itemize}
    \item \textbf{Role prompt}: The perturbation LLM acts as an expert in the specified programming language and is responsible for executing the required code perturbations.
    \item \textbf{Task prompt}: Describe the perturbation task which is divided into the following four parts: 
    
    \begin{itemize}
        \item \textbf{Prerequisites}: Specify the code snippet to be perturbed, providing the exact portion of the code where the perturbations should be applied.
        
        \item \textbf{Perturbation requirements}: Provide a description of the required perturbation and any boundary conditions, such as prohibiting the addition of default parameters unless the last parameter is already a default parameter.
        \item \textbf{Perturbation format}: Abstract the perturbation method into a template, making it easier for the perturbation LLM to understand and enabling it to generate more diverse perturbations based on predefined rules.
        \item \textbf{Example}: This part is optional. If more precise perturbations are needed, specific examples can be provided.
    \end{itemize}

    \item \textbf{Code prompt}: Contains all the code to be perturbed and the specified entry point for the perturbation function.
    \item \textbf{Answer prompt}: Instructs the perturbation LLM to return the modified code and entry point in JSON format.
\end{itemize}

We validate the intermediate perturbed code $c_{inter}$ to ensure it can be compiled and executed using our self-developed multi-language execution engine. For compiled languages such as C++, Rust, and GO, we use local compilers like \texttt{g++}, \texttt{cargo}, and \texttt{gc} to compile the code, respectively. Specifically, for Java, we verify the correct file name to prevent compilation failures caused by naming issues. For interpreted languages like Python, we use tools such as \texttt{pylint}~\cite{pylint} to check for syntax errors. Afterward, both the original and perturbed codes are evaluated by three LLMs (referred to as voter LLMs) to determine if they are semantically equivalent. The voter LLMs can only return \texttt{True} or \texttt{False}. The perturbation passes verification only if at least two voter LLMs return \texttt{True}. If $c_{inter}$ fails to compile or if fewer than two voter LLMs return \texttt{True}, the $c_{inter}$ does not pass verification. In this case, we do not provide the perturbation LLM with error feedback but instead let the perturbation LLM regenerate the perturbed code using a new perturbation method and $c_{pert}$.

\subsection{Perturbation method selection and optimization }\label{sec:method selection}

To apply appropriate perturbation methods to different code snippets and achieve improved results, we propose a method selection optimization algorithm, \alg{}, based on a genetic algorithm. In each iteration, the optimization algorithm aims to identify a more effective perturbation method for the code, which is then applied in the subsequent iteration. To avoid convergence to local optima, the algorithm introduces an element of randomness. The novel algorithm consists of three steps: similarity score initialization, similarity score iteration, and perturbation method selection. The details of similarity score iteration and perturbation method selection are provided in \autoref{alg:optimization-algorithm}. Note that, for clarity, we assume $c_{inter}$ always passes verification (see \autoref{sec:promt-generation}) after each perturbation, therefore, the verification process is omitted in \autoref{alg:optimization-algorithm}.

\subsubsection{Similarity score initialization}\label{sec:method/init}

At the start of the iteration, the initial overall similarity score $ss$ and minimum similarity score $mss$ are both set to 1. \alg{} randomly selects a perturbation method from each of the six method categories shown in \autoref{tab:all-methods} to iteratively perturb the code $c_{pert}$ sequentially, six times in total. In one iteration of the initialization phase, If $c_{inter}$ does not pass verification, then $og = 0$. Otherwise, as shown in lines 7–15 of \autoref{alg:optimization-algorithm}, we calculate $ss$. If $mss \ge ss$, \alg{} sets $c_{pert} = c_{inter}$, $og = mss - ss$, and $mss = ss$. Otherwise, $og$ is set to 0, with no changes to $mss$ or $c_{pert}$. 
Finally, $og_i$ is assigned the value of $og$. The larger the $og_i$, the more effective the perturbation, and the higher the probability that the method category $mc_i$ will be selected after initialization. Upon completion of the initialization phase, the initial $og_i$ values for each category are obtained as elements in $OG$.

\subsubsection{Similarity score iteration} The method for calculating the similarity scores has a significant impact on the effectiveness of the algorithm~\cite{liu2024eatvul}. This score needs to accurately reflect the differences between the perturbed code and the original code. Inspired by Riddell \ea{}~\cite{riddell2024quantifying}, we quantify the differences between the two code versions using both surface-level similarity and semantic similarity. 

Surface-level similarity represents the edit distance between two code strings. However, relying solely on surface-level similarity is insufficient, as perturbations like variable renaming do not alter the code structure. Semantic similarity, on the other hand, detects semantic differences between code versions and quantifies similarity accordingly. By combining these two types of scores to calculate the overall similarity score, \alg{} can more comprehensively select appropriate perturbation methods during iterations.

The calculation of the $ss$ is shown in \autoref{eq:fs}:
\begin{equation}
    ss = \mu \cdot s_1 + \nu \cdot s_2
    \label{eq:fs}
\end{equation}

\noindent where $ss, s_1, s_2, \mu, \nu \in [0,1]$ and $\mu + \nu = 1$. Here, $ss$ represents the overall similarity score, $s_1$ represents the surface-level similarity score, and $s_2$ represents the semantic similarity score. $\mu$ and $\nu$ are the respective weights. By adjusting the values of $\mu$ and $\nu$, \alg{} can prioritize methods that have a greater impact on either semantics or surface-level similarity.

The $mss$ represents the minimum similarity score during the iteration process. As in the initialization phase, if $ss$ is greater than $mss$, this indicates that the perturbation has increased code similarity, so the perturbation is discarded, and $og$ is set to 0. Otherwise, the perturbation is retained, with the corresponding $og = mss - ss$, and $mss$ is updated to $ss$. Even when $mss$ equals $ss$, we still retain this perturbation, as the cumulative effect of such perturbations is expected to be beneficial in subsequent iterations.

\subsubsection{Perturbation method selection}
To prevent \alg{} from repeatedly selecting a limited set of perturbation method categories, we use Boltzmann selection, where each category has a probability of being chosen. The Boltzmann selection algorithm is shown in \autoref{eq:bls}.
\begin{equation}\label{eq:bls}
    P_i = \frac{\exp\left(\frac{og_i}{T}\right)}{\sum_{j=1}^{N} \exp\left(\frac{og_j}{T}\right)}
\end{equation}

\noindent where $P_i$ represents the probability of selecting method category $mc_i$, and $og_i$ represents the optimization gain of each method category. $T$ is the temperature parameter, the larger the value of $T$, the more random the selection process. $N$ represents the total number of categories. Note that Boltzmann selection follows the principle that the smaller the $og_i$, the lower the probability that the corresponding category will be selected. Next, we randomly choose a perturbation method from that category to apply in the next iteration.

We now explain this process with a concrete example. As illustrated in \autoref{fig:order_of_perturbation}, we assume that the optimization gain for the \textit{\textbf{Condition Methods}} category is $og_{cm} = 0.2$. After applying the \textit{Div composed if} method, if the similarity between the code in the middle and the original code on the left decreases by 0.3 compared to $mss$, $og_{cm}$ is updated to 0.3, thereby increasing the probability of selecting the \textit{Div if else} method. Conversely, if a method from the \textit{\textbf{Logic Methods}} category (with $og_{lm} = 0.2$) is used to perturb the code on the right in \autoref{fig:code_trans_sample}, and the resulting similarity change is no more than 0.2 or even zero, $og_{lm}$ decreases, reducing the probability $P_{lm}$ of selecting a method from the \textit{\textbf{Logic Methods}} category. However, even if $og_{lm} = 0$, the probability $P_{lm}$ calculated by \autoref{eq:bls} remains greater than zero, albeit small. This ensures that \alg{} continues to explore the entire perturbation method space, while methods with higher perturbation effectiveness are selected with a greater probability.

In summary, \alg{} enhances this exploration process by selecting the more effective perturbation methods with higher probability in each iteration while allowing less effective methods to still be chosen at a lower probability. This approach promotes diversity in the perturbation method space and prevents the algorithm from becoming trapped in local optima, enabling the generation of complex perturbation combinations.

\subsection{Termination Conditions}
\label{sec:termination}
As shown in \autoref{alg:optimization-algorithm}, the conditions for ending the perturbation are twofold: \textit{(a)} the maximum number of iterations $maxIter$ is reached, and \textit{(b)} the $ss$ does not exceed the threshold $ss_{threshold}$. If either of these conditions is met, the iteration process terminates, yielding the final perturbed code.

\subsection{Implementation}
We now detail the implementation of \tool{}. GPT-4o is chosen for code perturbation due to its strong performance in both natural language and coding tasks, as well as its superior ability to interpret prompts. For the voting process, we selected more cost-effective LLMs, including GPT-3.5-turbo, DeepSeek-coder, and Claude 3.5-sonnet, to manage expenses while maintaining accuracy.
All of the LLMs used in \autoref{sec:approach} are accessed via paid APIs, as none of them run locally. Based on observations from the preliminary study, the cost of a single perturbation is approximately \$0.20 per instance.

\section{Evaluation}
\label{sec:RQ}

For the evaluation process, we aim to seek to address the following key research questions (RQs):

\begin{itemize}
    \item \textbf{RQ1:} How effective is \tool{} at perturbing code across multiple programming languages?
    \item \textbf{RQ2:} Does \alg{} successfully generate more effective combinations and orders of perturbations?
\end{itemize}

\subsection{Experimental Design}

\begin{figure}[htbp]  
    \centering     
    \includegraphics[width=\linewidth]{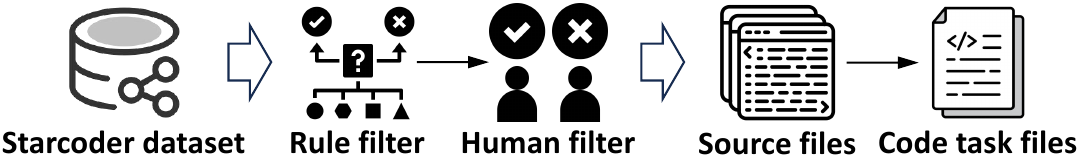}  
    \caption{Benchmark construction process}  
    \label{fig:benchmark}  
\end{figure}

As mentioned in \autoref{sec:pre-study}, to ensure the rigor of the experiments, we selected StarCoder as the LLM for evaluation and constructed the benchmark using The Stack dataset. Specifically, as shown in \autoref{fig:benchmark}, we begin by applying rule-based filtering and manual selection to The Stack to extract the source files. This is followed by the manual construction of a benchmark specifically designed for code completion tasks. These two phases are detailed in \autoref{sec:data-filter} and \autoref{sec:bench-construct} respectively.

To answer RQ1, we organize the experiments into five groups, each corresponding to a different programming language: C/C++, Python, Java, Rust, and Go. Due to the high cost associated with LLM API usage in \tool{}, we limit the selection to 20 code samples per group, culminating in a total of 100 original code samples 
from The Stack.

To answer RQ2, we include two additional groups of code samples to compare the effectiveness of random perturbation against the experimental group that utilized \alg{}. This approach ensures comprehensive evaluation and comparative analysis across different perturbation strategies.

\subsubsection{Data Filtering}\label{sec:data-filter}

The first step in constructing a benchmark from The Stack dataset involves data filtering, which includes both rule-based filtering and manual selection. We now detail these two parts.

First, we filter out the most unsuitable data using rule-based criteria. In this process, we exclude code files based on the following criteria: \textbf{(1) Insufficient lines} and \textbf{(2) Insufficient characters}. Specifically, we filter out code with fewer than 40 lines or fewer than 100 characters, as we observed that the repo-level code in benchmarks like CrossCodeEval~\cite{ding2024crosscodeeval} generally meets or exceeds these thresholds. While minimal code may simplify the perturbation process, it does not represent typical real-world scenarios and fails to demonstrate the applicability of \tool{} effectively. \textbf{(3) Uncompilable code}. We use a self-developed multi-language execution engine, described in \autoref{sec:promt-generation}, to verify code compilability. For compiled languages such as C/C++, Go, and Rust, we use \texttt{Tree-sitter}~\cite{treesitter} to check for the presence of a main function, adding an empty one if it is absent to ensure the code compiles without related errors. For Java, we verify the correct file name to prevent compilation failures caused by naming issues. For Python, we use the tool named \texttt{pylint}~\cite{pylint} to check for syntax errors without executing the program.

Since these rules cannot filter out all ineffective code files, such as those containing only class, method, or constant definitions, manual selection is also necessary. During manual screening, we specifically choose code with cross-file dependencies, multiple method definitions, or complex control flows. Ultimately, due to the substantial cost associated with using LLM APIs in \tool{}, we select a total of 100 source files across five languages, forming five distinct groups.

\subsubsection{Code Task Design}\label{sec:bench-construct}

 then proceed to design the code tasks for each group. 
After obtaining the source code, we use \tool{} to systematically perturb them and generate the corresponding perturbed code with the desired variations. We then proceed to carefully design the code tasks for each group to evaluate their performance.

Since The Stack is a training dataset that primarily consists of code without unit tests, descriptions, or related content, we opted to construct code completion tasks, as these are well-suited for such data. To evaluate the performance of \tool{} and \alg{} effectively within the limited code samples from The Stack, we constructed three types of code completion tasks for each group, involving the completion of 1, 3, and 5 lines of code. This approach resulted in a total of 600 tasks across the five groups: 300 tasks for the original code and 300 for the perturbed code. To assess the impact of code perturbations on LLM code completion, we marked sections only within the perturbed functions. For consistency, the same code snippets or logic were marked in both the original and perturbed versions. Furthermore, we selected two more groups of source code in Python and Go, applying randomly chosen perturbation methods to validate the effectiveness of \alg{}. The benchmark construction for these groups followed the same process, with 60 tasks per random perturbed group, totaling 120 tasks.
In total, we executed 720 code completion tasks using StarCoder. The experimental results obtained from these tasks serve as the answer to RQ1 and RQ2.

\subsubsection{Evaluation Metrics} 
For \tool{}, since our benchmark is constructed from The Stack, which primarily consists of code without unit tests, we cannot use unit test pass rates as an evaluation metric. CrossCodeEval, proposed by Ding \ea{}~\cite{ding2024crosscodeeval}, also lacks unit tests and instead uses \textit{code match} and \textit{identifier match} to measure the accuracy of key elements in the code. However, these metrics do not fully capture whether the code completed by LLMs is functionally equivalent to the original. To address this, we opted for a human assessment to determine if the completed code performs the same functionality as the original code, using accuracy as the evaluation metric for \tool{}. This metric reflects the number of code samples that are fully functionally equivalent.
Additionally, since \alg{} is designed to select optimal perturbation combinations, we use code similarity and accuracy as evaluation metrics to compare its effectiveness against the effectiveness of random perturbations.

\subsection{Experimental Setup}

We now define the LLMs used in the experiments, the selection of code similarity algorithms, and the hyperparameters employed in \tool{} and \alg{}. Additionally, we provide details on the human verification process and outline the experimental platform used for evaluation.

\noindent\textbf{Model} As mentioned before, we select StarCoder with 15.5B parameters as the evaluation model and set the maximum input token limit to 5k to ensure stable outputs. The perturbation LLM used is GPT-4, while the voter LLMs are GPT-3.5-turbo, DeepSeek-coder, and Claude 3.5-sonnet.

\noindent\textbf{Similarity Algorithm}  To comprehensively evaluate code similarity, we assess it from both surface-level and semantic perspectives. Similar to Riddell \ea{}~\cite{riddell2024quantifying}, we use the Levenshtein similarity score~\cite{levenshtein1966binary} for surface-level similarity, which is widely used due to its ease of computation. Semantic similarity is calculated using the \texttt{JPlag}~\cite{jplag} tool, a robust multi-language code plagiarism detector ideal for assessing similarity between perturbed code and original code.

\noindent\textbf{Hyperparameters} For hyperparameters in \tool{} and \alg{}, we set $\mu$ and $\nu$ in \autoref{eq:fs} to 0.5 to balance surface-level and semantic similarity. The temperature for Boltzmann selection is set to $T = 2$ to broaden exploration and avoid local optima. The iteration limit is 15, which is sufficient for identifying more effective perturbations. The $ss_{threshold}$ for early stopping is set at 0.2, ensuring that once this level is reached, the process halts because we have consistently observed through experimentation that further perturbations are unlikely to significantly reduce the similarity, making additional modifications redundant and inefficient.

\noindent\textbf{Human Verification} Since LLM hallucinations may lead to misjudgments, we manually audited all perturbed code for semantic consistency with the original. Inconsistent code was discarded and re-perturbed by perturbation LLM.

\noindent\textbf{Experiment Platform} The evaluation experiments are conducted on two NVIDIA A100 GPUs with 80GB of memory each, utilizing a Linux environment.

\section{Results}

In this section, we present the experimental results to answer the RQs proposed in \autoref{sec:RQ}.

\subsection{RQ1: How effective is \tool{} at perturbing code across multiple programming languages?}

\tool{} is designed as a cross-language code perturbation tool aimed at mitigating data contamination without altering the original functionality of the code. To answer RQ1, we evaluate \tool{} from four perspectives: code decontamination effectiveness, perturbation performance on multiple languages, accuracy on code completion tasks, and the correctness of the perturbed code.

\begin{figure*}[htbp]  
    \centering     
    \includegraphics[width=0.8\linewidth]{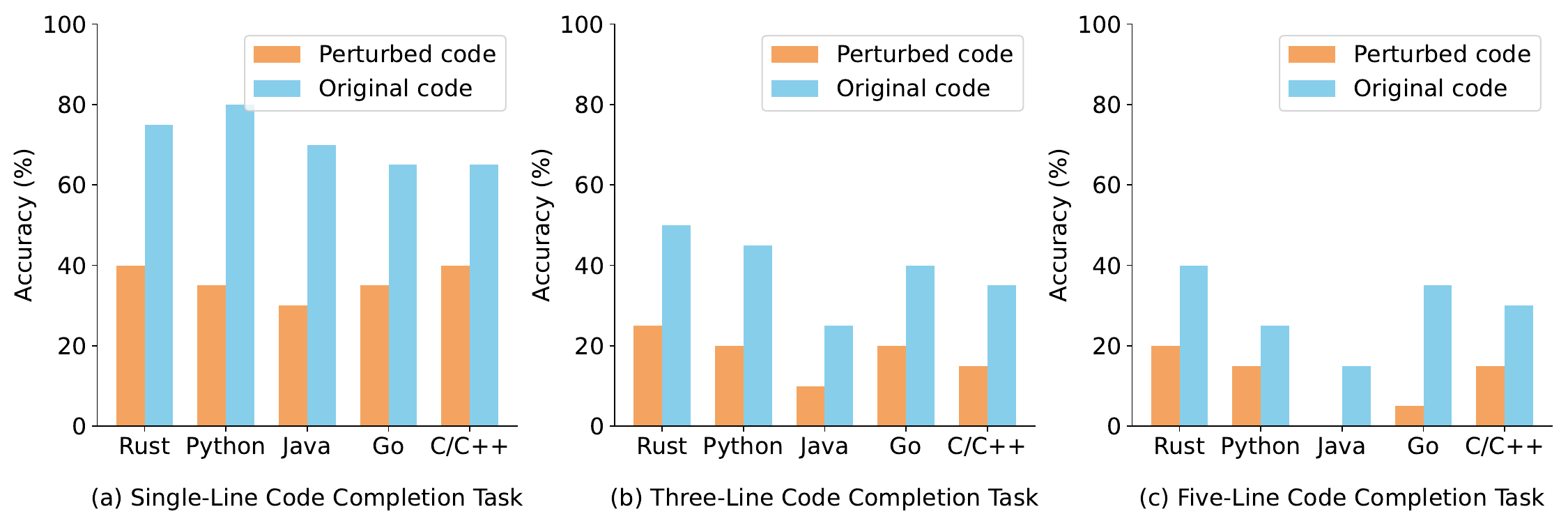}  
    \caption{Accuracy distribution across different code completion tasks for five languages}  
    \label{fig:five_lang_correct_rate}  
\end{figure*}

\noindent\textbf{Effect of Decontamination} As shown in \autoref{fig:five_lang_correct_rate}, the accuracy of code completion on perturbed code shows a significant decrease compared to the original, with an average reduction of \textbf{24.67\%}. The largest difference is observed in single-line completions, where the accuracy for perturbed Python code is \textbf{45\%} lower than for the original. The possible reason could be that masking only one line allows the original code to retain a high degree of integrity, enabling StarCoder to achieve higher accuracy. In contrast, the perturbed code exhibits reduced similarity to the original, making it difficult for StarCoder to rely on memorized patterns, resulting in a significant decline in accuracy. As the number of masked lines increases, the difficulty of code completion rises for both original and perturbed code, leading to further reductions in accuracy. Notably, the accuracy for perturbed code consistently remains lower than for the original. These results indicate that \tool{} effectively mitigates data leakage.

\begin{figure}[htbp]  
    \centering     
    \includegraphics[width=0.85\linewidth]{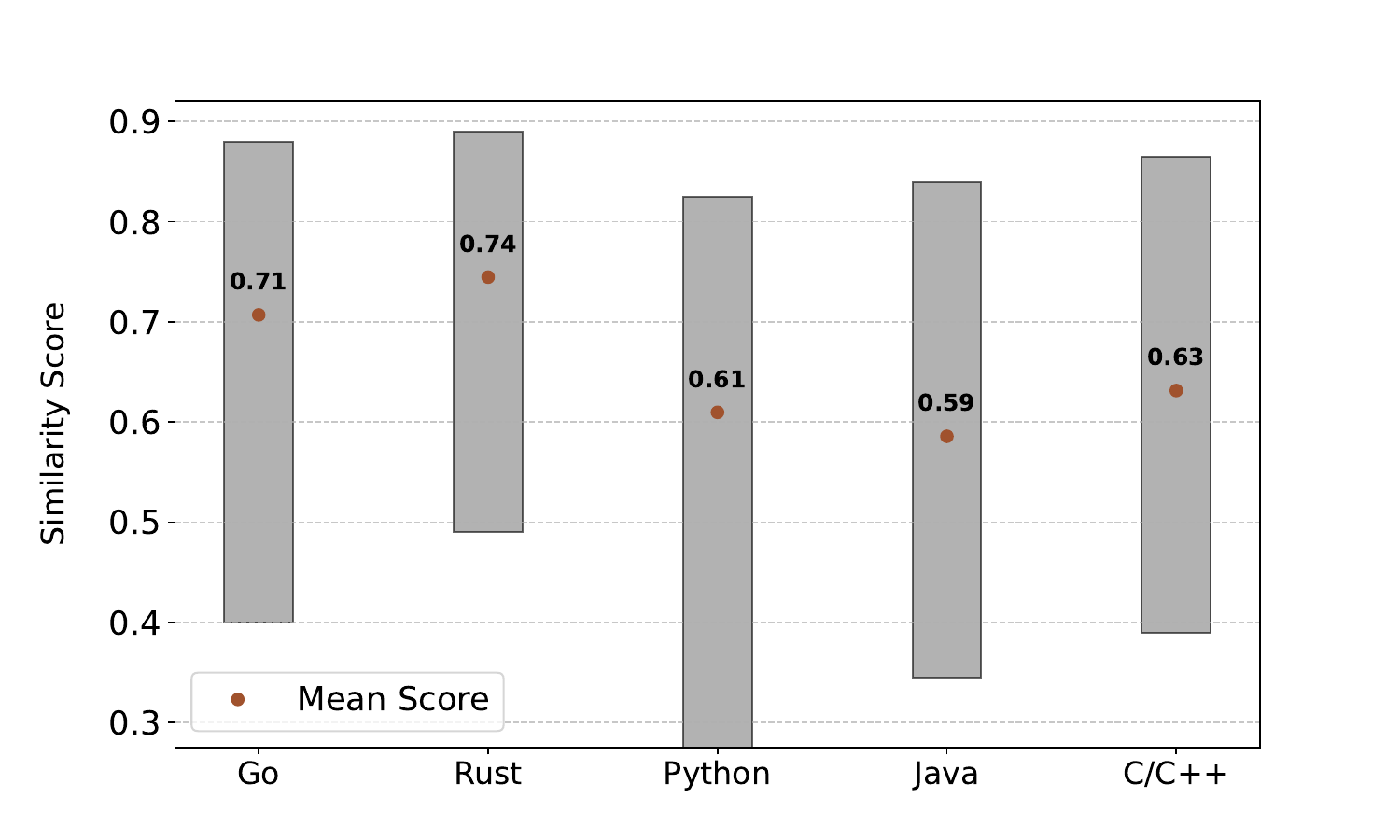}  
    \caption{Distribution interval plot of overall similarity scores}  
    \label{fig:similarity-scope}
\end{figure}

\noindent\textbf{Effect on Different Languages} \autoref{fig:similarity-scope} presents overall similarity scores across different languages, where \texttt{Mean Score} represents the average overall similarity score. The figure shows that all five languages experience varying degrees of reduction in overall similarity scores. Among them, Java, Python, and C/C++ demonstrate stronger perturbation effects, with average overall similarity scores around 0.6, proving that \tool{} effectively reduces code similarity scores. The higher similarity scores for Rust and Go can be attributed to their extensive standard libraries, which introduce fixed structures and logic, making them less susceptible to further perturbation. Nonetheless, as shown in \autoref{fig:five_lang_correct_rate}, the code completion accuracy for the perturbed Rust and Go code remains comparable to that of other programming languages, indicating that our perturbation methods are broadly applicable and effective across diverse programming languages.

\begin{figure}[htbp]  
    \centering     
    \includegraphics[width=0.9\linewidth]{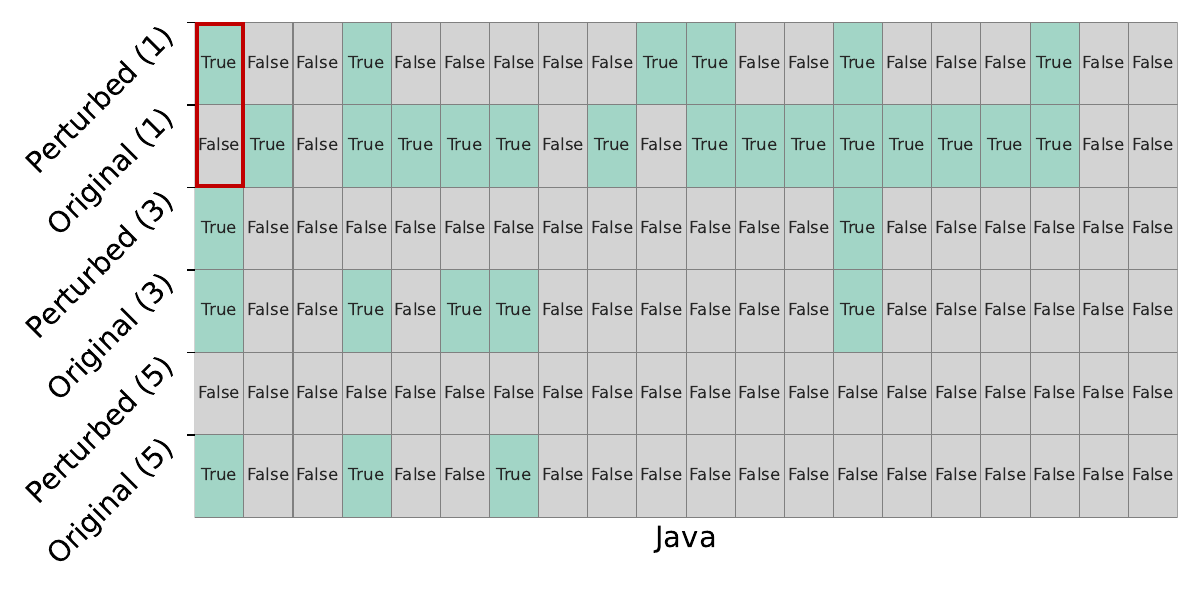}  
    \caption{Experimental results for Java code completion tasks}  
    \label{fig:correct_hot_map}

\end{figure}

\noindent\textbf{Effect on Code Completion} An interesting observation is that code perturbation does not always reduce code completion accuracy, aligning with the findings of Cao \ea{}~\cite{cao2024concerned}. \autoref{fig:correct_hot_map} presents the results of Java group code completion tasks, with gray indicating failure and green indicating success. \texttt{Perturbed (1)} and \texttt{Original (1)} represent the single-line code completion tasks for perturbed and original code, respectively, and the same applies to other cases. For instance, in the results marked by the red box in the figure, the perturbed code's completion is successful, while the original code's completion fails. This phenomenon is observed in other groups as well. The reason for this may be that the perturbed code introduces hints that lower the difficulty of the completion task. For example, as shown in the left code in \autoref{fig:code_trans_sample}, the \emph{Extract if} method encapsulates complex \texttt{if} statements within a function. If StarCoder is asked to complete a complex statement such as \verb| if  a < b  or  c < d| in the original code, it faces a greater challenge, whereas in the perturbed code, it only needs to identify which function to call, making the task easier. However, despite such occurrences, the overall accuracy for perturbed code remains significantly lower than for the original code, indicating that \tool{} effectively utilizes code perturbations to mitigate data contamination.

\noindent\textbf{Correctness of Perturbed Code} After manual review, the number of incorrect perturbed code samples per language was as follows: four errors in Java, and between 0 and 3 errors in the remaining languages. For incorrect samples, reapplying perturbation generally resolved semantic inconsistencies with high probability, typically requiring \textbf{no more than} two attempts. This demonstrates that \tool{} generates perturbed code with high accuracy and low correction cost.

\noindent\textbf{Summary} From the above analysis, we can conclude that \tool{} effectively mitigates contamination across benchmarks for different languages. Although some code may be challenging to perturb due to fixed structures, and certain perturbation methods may unintentionally introduce hints that reduce the difficulty of code completion tasks, \tool{} minimizes or eliminates these effects, achieving the goal of decontamination. Furthermore, the perturbed code generated by \tool{} remains largely semantically consistent with the original, significantly reducing the need for manual inspection.

\subsection{RQ2: Does \alg{} successfully generate more effective combinations and orders of perturbations?}

To address RQ2, we evaluate \alg{} from two perspectives: code similarity and the accuracy of code completion tasks.

In terms of accuracy, as shown in \autoref{tab:correct_random_selection}, \texttt{Single} represents single-line code completion tasks, with \texttt{Three} and \texttt{Five} denoting their respective counterparts. The findings show that random perturbations yield higher accuracy in code completion, while \alg{} results in an average accuracy reduction of \textbf{15\%} and a maximum reduction of \textbf{25\%}. The random algorithm shows a smaller accuracy decline compared to the original code, with an average reduction of \textbf{11.67\%}, whereas the \alg{} group displays an average reduction of \textbf{26.67\%}.

In terms of code similarity, as illustrated in \autoref{tab:random_selection}, we present surface-level similarity, semantic similarity, and the overall similarity score $ss$ for Python and Go. Across these metrics, \alg{} consistently achieves lower similarity scores than the random algorithm, showing an average reduction in $ss$ of \textbf{7.01\%} and a maximum reduction of \textbf{42.86\%}. These results, combined with the reduction in accuracy, suggest that code similarity can effectively indicate decontamination and that reducing similarity is a practical strategy for achieving it.

However, the random algorithm is not always less effective than \alg{}. In some cases, it achieves lower similarity scores, indicating that random perturbations can sometimes produce better combinations. This occurs because \alg{} aims to find a better solution rather than an optimal one. Note that, we introduce certain elements of randomness into the method selection process, such as Boltzmann selection and the random choice of methods within specific categories. The introduction of randomness can occasionally lead to less effective perturbation methods, which may reduce the overall perturbation effect. However, this strategy helps avoid \alg{} falling into local optima, making it an acceptable trade-off.

\noindent\textbf{Summary} \alg{} effectively identifies superior code perturbation methods. Additionally, code similarity can be used as a measure of decontamination, with lowering similarity proving to be an effective approach to achieve it.

\begin{table}[]
\caption{Accuracy comparison between \alg{} and random perturbation}
\label{tab:correct_random_selection}
\center
\resizebox{0.8\linewidth}{!}{%
\begin{tabular}{|c|ccc|ccc|}
\hline
              & \multicolumn{3}{c|}{Python}                                     & \multicolumn{3}{c|}{Go}                                         \\ \cline{2-7} 
              & \multicolumn{1}{c|}{Single} & \multicolumn{1}{c|}{Three} & Five & \multicolumn{1}{c|}{Single} & \multicolumn{1}{c|}{Three} & Five \\ \hline
\alg          & \multicolumn{1}{c|}{35\%}   & \multicolumn{1}{c|}{20\%}  & 15\% & \multicolumn{1}{c|}{35\%}   & \multicolumn{1}{c|}{20\%}  & 5\%  \\ \hline
Random        & \multicolumn{1}{c|}{60\%}   & \multicolumn{1}{c|}{30\%}  & 20\% & \multicolumn{1}{c|}{60\%}   & \multicolumn{1}{c|}{25\%}  & 25\% \\ \hline
Original code & \multicolumn{1}{c|}{80\%}   & \multicolumn{1}{c|}{45\%}  & 25\% & \multicolumn{1}{c|}{65\%}   & \multicolumn{1}{c|}{40\%}  & 35\% \\ \hline
\end{tabular}%
}

\end{table}

\begin{table}[]
\caption{Comparison of code similarity scores between \alg{} and random perturbation}
\label{tab:random_selection}
\resizebox{\linewidth}{!}{%
\begin{tabular}{|c|cccccc|cccccc|}
\hline
\multirow{3}{*}{\begin{tabular}[c]{@{}c@{}}Code \\ ID\end{tabular}} & \multicolumn{6}{c|}{Python}                                                                                                                                                                                                                  & \multicolumn{6}{c|}{Go}                                                                                                                                                                                                                      \\ \cline{2-13} 
                                                                    & \multicolumn{2}{c|}{SuS}                                                            & \multicolumn{2}{c|}{SeS}                                                            & \multicolumn{2}{c|}{SS}                                          & \multicolumn{2}{c|}{SuS}                                                            & \multicolumn{2}{c|}{SeS}                                                            & \multicolumn{2}{c|}{SS}                                          \\ \cline{2-13} 
                                                                    & \multicolumn{1}{c|}{M}                   & \multicolumn{1}{c|}{R}                   & \multicolumn{1}{c|}{M}                   & \multicolumn{1}{c|}{R}                   & \multicolumn{1}{c|}{M}                    & R                    & \multicolumn{1}{c|}{M}                   & \multicolumn{1}{c|}{R}                   & \multicolumn{1}{c|}{M}                   & \multicolumn{1}{c|}{R}                   & \multicolumn{1}{c|}{M}                    & R                    \\ \hline
1                                                                   & \multicolumn{1}{c|}{0.80}                & \multicolumn{1}{c|}{{\ul \textbf{0.81}}} & \multicolumn{1}{c|}{0.73}                & \multicolumn{1}{c|}{{\ul \textbf{0.75}}} & \multicolumn{1}{c|}{0.765}                & {\ul \textbf{0.78}}  & \multicolumn{1}{c|}{0.72}                & \multicolumn{1}{c|}{{\ul \textbf{0.76}}} & \multicolumn{1}{c|}{0.82}                & \multicolumn{1}{c|}{{\ul \textbf{0.92}}} & \multicolumn{1}{c|}{0.77}                 & {\ul \textbf{0.84}}  \\ \hline
2                                                                   & \multicolumn{1}{c|}{0.55}                & \multicolumn{1}{c|}{{\ul \textbf{0.77}}} & \multicolumn{1}{c|}{0.27}                & \multicolumn{1}{c|}{{\ul \textbf{0.41}}} & \multicolumn{1}{c|}{0.41}                 & {\ul \textbf{0.59}}  & \multicolumn{1}{c|}{0.75}                & \multicolumn{1}{c|}{{\ul \textbf{0.78}}} & \multicolumn{1}{c|}{0.95}                & \multicolumn{1}{c|}{{\ul \textbf{0.99}}} & \multicolumn{1}{c|}{0.85}                 & {\ul \textbf{0.885}} \\ \hline
3                                                                   & \multicolumn{1}{c|}{0.70}                & \multicolumn{1}{c|}{0.70}                & \multicolumn{1}{c|}{{\ul \textbf{0.78}}} & \multicolumn{1}{c|}{0.39}                & \multicolumn{1}{c|}{{\ul \textbf{0.74}}}  & 0.545                & \multicolumn{1}{c|}{{\ul \textbf{0.51}}} & \multicolumn{1}{c|}{0.48}                & \multicolumn{1}{c|}{0.78}                & \multicolumn{1}{c|}{{\ul \textbf{0.84}}} & \multicolumn{1}{c|}{0.645}                & {\ul \textbf{0.660}} \\ \hline
4                                                                   & \multicolumn{1}{c|}{0.66}                & \multicolumn{1}{c|}{{\ul \textbf{0.69}}} & \multicolumn{1}{c|}{0.69}                & \multicolumn{1}{c|}{0.69}                & \multicolumn{1}{c|}{0.675}                & {\ul \textbf{0.69}}  & \multicolumn{1}{c|}{{\ul \textbf{0.81}}} & \multicolumn{1}{c|}{0.74}                & \multicolumn{1}{c|}{0.94}                & \multicolumn{1}{c|}{{\ul \textbf{0.98}}} & \multicolumn{1}{c|}{{\ul \textbf{0.875}}} & 0.86                 \\ \hline
5                                                                   & \multicolumn{1}{c|}{0.71}                & \multicolumn{1}{c|}{{\ul \textbf{0.86}}} & \multicolumn{1}{c|}{0.91}                & \multicolumn{1}{c|}{{\ul \textbf{0.96}}} & \multicolumn{1}{c|}{0.81}                 & {\ul \textbf{0.91}}  & \multicolumn{1}{c|}{{\ul \textbf{0.82}}} & \multicolumn{1}{c|}{0.81}                & \multicolumn{1}{c|}{0.89}                & \multicolumn{1}{c|}{{\ul \textbf{0.95}}} & \multicolumn{1}{c|}{0.855}                & {\ul \textbf{0.88}}  \\ \hline
6                                                                   & \multicolumn{1}{c|}{{\ul \textbf{0.71}}} & \multicolumn{1}{c|}{0.68}                & \multicolumn{1}{c|}{0.37}                & \multicolumn{1}{c|}{{\ul \textbf{0.83}}} & \multicolumn{1}{c|}{0.54}                 & {\ul \textbf{0.755}} & \multicolumn{1}{c|}{{\ul \textbf{0.41}}} & \multicolumn{1}{c|}{0.38}                & \multicolumn{1}{c|}{{\ul \textbf{0.56}}} & \multicolumn{1}{c|}{0.48}                & \multicolumn{1}{c|}{{\ul \textbf{0.485}}} & 0.43                 \\ \hline
7                                                                   & \multicolumn{1}{c|}{0.67}                & \multicolumn{1}{c|}{{\ul \textbf{0.71}}} & \multicolumn{1}{c|}{0.53}                & \multicolumn{1}{c|}{{\ul \textbf{0.72}}} & \multicolumn{1}{c|}{0.60}                 & {\ul \textbf{0.715}} & \multicolumn{1}{c|}{0.48}                & \multicolumn{1}{c|}{{\ul \textbf{0.73}}} & \multicolumn{1}{c|}{0.32}                & \multicolumn{1}{c|}{{\ul \textbf{0.51}}} & \multicolumn{1}{c|}{0.40}                 & {\ul \textbf{0.62}}  \\ \hline
8                                                                   & \multicolumn{1}{c|}{0.82}                & \multicolumn{1}{c|}{{\ul \textbf{0.83}}} & \multicolumn{1}{c|}{0.83}                & \multicolumn{1}{c|}{{\ul \textbf{0.90}}} & \multicolumn{1}{c|}{0.825}                & {\ul \textbf{0.865}} & \multicolumn{1}{c|}{{\ul \textbf{0.64}}} & \multicolumn{1}{c|}{0.51}                & \multicolumn{1}{c|}{0.87}                & \multicolumn{1}{c|}{0.87}                & \multicolumn{1}{c|}{{\ul \textbf{0.755}}} & 0.69                 \\ \hline
9                                                                   & \multicolumn{1}{c|}{0.71}                & \multicolumn{1}{c|}{{\ul \textbf{0.82}}} & \multicolumn{1}{c|}{0.53}                & \multicolumn{1}{c|}{{\ul \textbf{0.76}}} & \multicolumn{1}{c|}{0.62}                 & {\ul \textbf{0.79}}  & \multicolumn{1}{c|}{0.68}                & \multicolumn{1}{c|}{{\ul \textbf{0.80}}} & \multicolumn{1}{c|}{{\ul \textbf{0.81}}} & \multicolumn{1}{c|}{0.63}                & \multicolumn{1}{c|}{0.745}                & {\ul \textbf{0.865}} \\ \hline
10                                                                  & \multicolumn{1}{c|}{0.77}                & \multicolumn{1}{c|}{{\ul \textbf{0.83}}} & \multicolumn{1}{c|}{0.71}                & \multicolumn{1}{c|}{{\ul \textbf{0.72}}} & \multicolumn{1}{c|}{0.74}                 & {\ul \textbf{0.775}} & \multicolumn{1}{c|}{{\ul \textbf{0.60}}} & \multicolumn{1}{c|}{0.59}                & \multicolumn{1}{c|}{0.82}                & \multicolumn{1}{c|}{{\ul \textbf{0.89}}} & \multicolumn{1}{c|}{0.71}                 & {\ul \textbf{0.74}}  \\ \hline
11                                                                  & \multicolumn{1}{c|}{{\ul \textbf{0.75}}} & \multicolumn{1}{c|}{0.73}                & \multicolumn{1}{c|}{0.75}                & \multicolumn{1}{c|}{{\ul \textbf{0.88}}} & \multicolumn{1}{c|}{0.75}                 & {\ul \textbf{0.805}} & \multicolumn{1}{c|}{{\ul \textbf{0.59}}} & \multicolumn{1}{c|}{0.44}                & \multicolumn{1}{c|}{{\ul \textbf{0.84}}} & \multicolumn{1}{c|}{0.52}                & \multicolumn{1}{c|}{{\ul \textbf{0.715}}} & 0.48                 \\ \hline
12                                                                  & \multicolumn{1}{c|}{0.82}                & \multicolumn{1}{c|}{{\ul \textbf{0.83}}} & \multicolumn{1}{c|}{{\ul \textbf{0.96}}} & \multicolumn{1}{c|}{0.88}                & \multicolumn{1}{c|}{{\ul \textbf{0.89}}}  & 0.855                & \multicolumn{1}{c|}{0.73}                & \multicolumn{1}{c|}{{\ul \textbf{0.76}}} & \multicolumn{1}{c|}{0.94}                & \multicolumn{1}{c|}{{\ul \textbf{0.97}}} & \multicolumn{1}{c|}{0.835}                & {\ul \textbf{0.865}} \\ \hline
13                                                                  & \multicolumn{1}{c|}{0.49}                & \multicolumn{1}{c|}{{\ul \textbf{0.51}}} & \multicolumn{1}{c|}{{\ul \textbf{0.6}}}  & \multicolumn{1}{c|}{0.17}                & \multicolumn{1}{c|}{{\ul \textbf{0.545}}} & 0.34                 & \multicolumn{1}{c|}{0.65}                & \multicolumn{1}{c|}{{\ul \textbf{0.89}}} & \multicolumn{1}{c|}{0.75}                & \multicolumn{1}{c|}{{\ul \textbf{0.98}}} & \multicolumn{1}{c|}{0.70}                 & {\ul \textbf{0.935}} \\ \hline
14                                                                  & \multicolumn{1}{c|}{0.74}                & \multicolumn{1}{c|}{{\ul \textbf{0.86}}} & \multicolumn{1}{c|}{0.77}                & \multicolumn{1}{c|}{{\ul \textbf{0.92}}} & \multicolumn{1}{c|}{0.755}                & {\ul \textbf{0.89}}  & \multicolumn{1}{c|}{0.61}                & \multicolumn{1}{c|}{{\ul \textbf{0.78}}} & \multicolumn{1}{c|}{0.80}                & \multicolumn{1}{c|}{{\ul \textbf{0.85}}} & \multicolumn{1}{c|}{0.705}                & {\ul \textbf{0.815}} \\ \hline
15                                                                  & \multicolumn{1}{c|}{0.66}                & \multicolumn{1}{c|}{{\ul \textbf{0.80}}} & \multicolumn{1}{c|}{0.66}                & \multicolumn{1}{c|}{{\ul \textbf{0.84}}} & \multicolumn{1}{c|}{0.66}                 & {\ul \textbf{0.82}}  & \multicolumn{1}{c|}{0.52}                & \multicolumn{1}{c|}{{\ul \textbf{0.91}}} & \multicolumn{1}{c|}{0.83}                & \multicolumn{1}{c|}{{\ul \textbf{0.96}}} & \multicolumn{1}{c|}{0.675}                & {\ul \textbf{0.935}} \\ \hline
16                                                                  & \multicolumn{1}{c|}{0.62}                & \multicolumn{1}{c|}{{\ul \textbf{0.64}}} & \multicolumn{1}{c|}{{\ul \textbf{0.48}}} & \multicolumn{1}{c|}{0.47}                & \multicolumn{1}{c|}{0.55}                 & {\ul \textbf{0.555}} & \multicolumn{1}{c|}{0.85}                & \multicolumn{1}{c|}{{\ul \textbf{0.89}}} & \multicolumn{1}{c|}{0.91}                & \multicolumn{1}{c|}{{\ul \textbf{0.98}}} & \multicolumn{1}{c|}{0.88}                 & {\ul \textbf{0.935}} \\ \hline
17                                                                  & \multicolumn{1}{c|}{0.55}                & \multicolumn{1}{c|}{{\ul \textbf{0.77}}} & \multicolumn{1}{c|}{0.00}                & \multicolumn{1}{c|}{0.00}                & \multicolumn{1}{c|}{0.275}                & {\ul \textbf{0.385}} & \multicolumn{1}{c|}{0.72}                & \multicolumn{1}{c|}{{\ul \textbf{0.80}}} & \multicolumn{1}{c|}{{\ul \textbf{0.96}}} & \multicolumn{1}{c|}{0.85}                & \multicolumn{1}{c|}{{\ul \textbf{0.84}}}  & 0.825                \\ \hline
18                                                                  & \multicolumn{1}{c|}{0.59}                & \multicolumn{1}{c|}{{\ul \textbf{0.69}}} & \multicolumn{1}{c|}{0.24}                & \multicolumn{1}{c|}{{\ul \textbf{0.37}}} & \multicolumn{1}{c|}{0.415}                & {\ul \textbf{0.53}}  & \multicolumn{1}{c|}{0.40}                & \multicolumn{1}{c|}{{\ul \textbf{0.43}}} & \multicolumn{1}{c|}{{\ul \textbf{0.52}}} & \multicolumn{1}{c|}{0.21}                & \multicolumn{1}{c|}{{\ul \textbf{0.46}}}  & 0.32                 \\ \hline
19                                                                  & \multicolumn{1}{c|}{{\ul \textbf{0.69}}} & \multicolumn{1}{c|}{0.67}                & \multicolumn{1}{c|}{{\ul \textbf{0.45}}} & \multicolumn{1}{c|}{0.26}                & \multicolumn{1}{c|}{{\ul \textbf{0.57}}}  & 0.465                & \multicolumn{1}{c|}{0.64}                & \multicolumn{1}{c|}{{\ul \textbf{0.69}}} & \multicolumn{1}{c|}{0.43}                & \multicolumn{1}{c|}{{\ul \textbf{0.95}}} & \multicolumn{1}{c|}{0.535}                & {\ul \textbf{0.85}}  \\ \hline
20                                                                  & \multicolumn{1}{c|}{0.57}                & \multicolumn{1}{c|}{{\ul \textbf{0.76}}} & \multicolumn{1}{c|}{0.27}                & \multicolumn{1}{c|}{{\ul \textbf{0.71}}} & \multicolumn{1}{c|}{0.42}                 & {\ul \textbf{0.735}} & \multicolumn{1}{c|}{0.61}                & \multicolumn{1}{c|}{{\ul \textbf{0.81}}} & \multicolumn{1}{c|}{0.80}                & \multicolumn{1}{c|}{{\ul \textbf{0.89}}} & \multicolumn{1}{c|}{0.705}                & {\ul \textbf{0.85}}  \\ \hline
Total                                                               & \multicolumn{1}{c|}{3}                   & \multicolumn{1}{c|}{{\ul \textbf{16}}}   & \multicolumn{1}{c|}{5}                   & \multicolumn{1}{c|}{{\ul \textbf{13}}}   & \multicolumn{1}{c|}{4}                    & {\ul \textbf{16}}    & \multicolumn{1}{c|}{7}                   & \multicolumn{1}{c|}{{\ul \textbf{13}}}   & \multicolumn{1}{c|}{5}                   & \multicolumn{1}{c|}{{\ul \textbf{14}}}   & \multicolumn{1}{c|}{6}                    & {\ul \textbf{14}}    \\ \hline
\end{tabular}%
}

\vspace{0.1cm}

\footnotesize{* SuS denotes the surface-level similarity score, SeS represents the semantic similarity score, and SS stands for the overall similarity score $ss$. M refers to \alg{}, and R denotes random perturbation. \\
* Scores marked with ``\_'' indicate higher similarity scores between the two perturbation methods, while scores without a mark indicate no difference. The \texttt{Total} row tallies the counts of scores marked with ``\_''.}

\end{table}
\section{Limitation}

\noindent\textbf{Benchmark and Model} Since we constructed the benchmark based on the publicly available The Stack, tasks like data cleaning and code task construction required significant effort, resulting in a relatively small benchmark. Additionally, we did not find other publicly available training datasets paired with corresponding LLMs for code, limiting this study to an evaluation of StarCoder. In the future, we will attempt to evaluate the performance of \tool{} on more datasets.

\noindent\textbf{Code Task} The only code task used for testing the LLM is code completion. This limitation arises primarily because The Stack is a training dataset, lacking unit test cases, docstrings, and related information, making the construction of other code tasks challenging. In the future, tasks such as code classification could be employed to assess the decontamination effectiveness of \tool{} across different datasets.

\section{Related Work}
\subsection{Benchmark for Code LLM}
Recently, numerous datasets have been developed for evaluating the coding capabilities of LLMs. HumanEval-X~\cite{zheng2023codegeex} extends HumanEval~\cite{chen2021evaluating} to over programming languages, enabling assessments of LLMs' multilingual code completion abilities. Additionally, datasets tailored for specific tasks have been introduced, such as XCODEEVAL~\cite{khan2024xcodeeval}, BigCodeBench~\cite{zhuo2024bigcodebench}, and CoderUJB~\cite{zeng2024coderujb}. Among these, XCODEEVAL~\cite{khan2024xcodeeval} supports multiple programming languages, although its samples are not at the repository level. In contrast, BigCodeBench~\cite{zhuo2024bigcodebench} provides repository-level samples, as does Feng \ea{}~\cite{feng2024complexcodeeval} proposed ComplexCodeEval, which was designed for diverse downstream tasks. However, it requires further adjustments to be used for specific evaluation purposes.

Despite their utility, many of these benchmarks face challenges related to data leakage. This issue arises because they often derive data either directly from real-world sources or by improving upon existing datasets, making it difficult to avoid overlap with data used during LLM pre-training and fine-tuning. As a result, the evaluation scores may not fully reflect the models' generalization capabilities, highlighting the need for more robust contamination-free benchmarks.

\subsection{The Method of Decontamination}

Currently, a popular method for mitigating data contamination is to evaluate LLMs using datasets released after the LLMs. Khan \ea{}~\cite{khan2024xcodeeval} introduced timestamps in their constructed dataset, XcodeEval, to identify potential data contamination, a practice also adopted by~\cite{feng2024complexcodeeval}. In addition to incorporating timestamps into datasets, Silva \ea{}~\cite{silva2024repairbench}, White \ea{}~\cite{white2024livebench}, and Jain \ea{}~\cite{jain2024livecodebench} all continuously update their own benchmarks to mitigate data contamination. Ding \ea{}~\cite{ding2024crosscodeeval} constructed the CrossCodeEval dataset using code from repositories active between 2023-03-05 and 2023-06-15. However, even with timely updates, data still faces the risk of contamination due to outdated information~\cite{yang2023rethinking}. Moreover, researchers have employed data perturbation methods to mitigate contamination. Xia \ea{}~\cite{xia2024top} modified prompts based on HumanEval across five dimensions, thereby generating new benchmarks. Liu \ea{}~\cite{liu2024your} expanded the test cases of HumanEval, increasing the difficulty for LLMs to pass the tests. Zhu \ea{}~\cite{zhu2024dynamic}, inspired by psychometrics, reconstructed existing problems across three dimensions to dynamically assess LLMs. These methods have all consistently resulted in decreased performance for LLMs, clearly indicating their effectiveness in successfully mitigating contamination.

\subsection{Code Perturbation}

Code perturbation, as a data augmentation method, has wide applications in various scenarios. In the context of malicious attacks, Zhang \ea{}~\cite{zhang2024codebert} proposed a method for generating attack code using code perturbation to mislead models. Similarly, Yu \ea{}~\cite{yu2022data} found that minor perturbations can cause LLMs to make incorrect predictions. Yang \ea{}~\cite{yang2024exploiting} used semantics-preserving transformations to enable adversarial code to transfer more effectively across different models. Conversely, code perturbation can also be employed for data security purposes. For example, Xian \ea{}~\cite{xian2024transformcode} introduced a novel data augmentation technique using AST transformations to enhance contrastive learning. Jiang \ea{}~\cite{jiang2024detecting} employed equivalent expression transformations to detect bugs in database engines, based on the principle that query results in the database should remain consistent before and after transformation. Yang \ea{}~\cite{yang2024srcmarker} embedded watermarks in code through various perturbations to track code ownership and prevent malicious use. Wang \ea{}~\cite{wang2022recoderobustnessevaluationcode} developed a series of semantics-preserving code perturbation methods to construct a benchmark for evaluating code robustness. Similar to these works, we leverage the data augmentation capabilities of code perturbation to transform original code into forms that LLMs have not previously learned, thereby mitigating data contamination issues during the evaluation of LLM.

\section{Conclusion}

In this work, we propose \tool{} and \alg{} to mitigate data contamination issues in LLMs. \tool{} supports multiple languages and applies 26 different perturbation methods, while \alg{} selects the most effective combinations for each perturbation to enhance its impact. Experimental results demonstrate that \tool{} effectively mitigates data contamination issues, and that the perturbation effects of the \alg{} outperform those of random selection methods.

\newpage
\bibliographystyle{IEEEtranS}
\bibliography{main}

% \newpage
% \input{Sections/10.appendix}

\end{document}